\documentclass[pra,twocolumn,amsmath,amssymb,amsthm,superscriptaddress]{revtex4-2}

\usepackage{graphicx,amsmath,amsfonts,relsize,epstopdf,upgreek,color,mathtools,bm,times,braket,rotating,tabularx,booktabs,setspace,makecell}
\usepackage[colorlinks=true,linkcolor=blue,citecolor=blue,urlcolor =blue]{hyperref}
\usepackage[hyphenbreaks]{breakurl}
\usepackage[T1]{fontenc}

\newcommand{\rvec}[1]{\pmb{#1}}

\newcommand{\tr}[1]{\mathrm{tr}\left\{#1\right\}}
\newcommand{\ptr}[2]{\mathrm{tr}_{#1}\left\{#2\right\}}

\newcommand{\D}{\mathrm{d}}
\newcommand{\I}{\mathrm{i}}

\newcommand{\E}[1]{\mathrm{e}^{\mbox{\footnotesize$#1$}}}

\newcommand{\ML}{\widehat{\varrho}_\textsc{ml}}
\newcommand{\PR}{\mathrm{pr}}
\newcommand{\RB}{\mathrm{RB}}
\newcommand{\deff}{d_{\mathrm{eff}}}
\newcommand{\dRB}{d_{\mathrm{RB}}}

\newcommand{\vacket}{\ket{\textsc{vac}}}
\newcommand{\vacbra}{\bra{\textsc{vac}}}
\newcommand{\appropto}{\mathrel{\vcenter{
			\offinterlineskip\halign{\hfil$##$\cr
				\propto\cr\noalign{\kern2pt}\sim\cr\noalign{\kern-2pt}}}}}

\begin{document}
	
	\title{Relative-belief inference in quantum information theory}
	
	\author{Y. S. Teo}
	\email{ys\_teo@snu.ac.kr}
	\affiliation{Department of Physics and Astronomy, 
		Seoul National University, 08826 Seoul, South Korea}
	
	\author{S. U. Shringarpure}
	\email{saurabh.s@snu.ac.kr}
	\affiliation{Department of Physics and Astronomy, 
		Seoul National University, 08826 Seoul, South Korea}
	
	\author{H. Jeong}
	\email{h.jeong37@gmail.com}
	\affiliation{Department of Physics and Astronomy, 
		Seoul National University, 08826 Seoul, South Korea}
	
	\author{N. Prasannan}
	\affiliation{Integrated Quantum Optics Group, Applied Physics, University of Paderborn, 33098 Paderborn, Germany}
	
	\author{B. Brecht}
	\affiliation{Integrated Quantum Optics Group, Applied Physics, University of Paderborn, 33098 Paderborn, Germany}
	
	\author{C. Silberhorn}
	\affiliation{Integrated Quantum Optics Group, Applied Physics, University of Paderborn, 33098 Paderborn, Germany}
	
	\author{M. Evans}
	\email{mevansthree.evans@utoronto.ca}
	\affiliation{Department of Statistical Sciences, University of Toronto, Toronto, Ontario, M5S 3G3, Canada}
	
	\author{D. Mogilevtsev}
	\affiliation{B. I. Stepanov Institute of Physics, NAS of Belarus, Nezavisimosti ave. 68, 220072 Minsk, Belarus}
	
	\author{L. L. S{\'a}nchez-Soto}
	\affiliation{Departamento de {\'O}ptica, Facultad de F{\'i}sica, Universidad Complutense, 28040 Madrid, Spain}
	\affiliation{Max-Planck-Institut f{\"u}r die Physik des Lichts, Staudtstra{\ss}e 2, 91058 Erlangen, Germany}
	
	\begin{abstract}
		We introduce the framework of Bayesian relative belief that directly evaluates whether or not the experimental data at hand supports a given hypothesis regarding a quantum system by directly comparing the prior and posterior probabilities for the hypothesis. In model-dimension certification tasks, we show that the relative belief procedure typically chooses Hilbert spaces that are never smaller in dimension than those selected from optimizing a broad class of information criteria, including Akaike's criterion. As a concrete and focused exposition of this powerful evidence-based technique, we apply the relative belief procedure to an important application: \emph{state reconstruction of imperfect quantum sources}. In particular, just by comparing prior and posterior probabilities based on data, we demonstrate its capability of tracking multiphoton emissions using (realistically lossy) single-photon detectors in order to assess the actual quality of photon sources without making \emph{ad hoc} assumptions, thereby reliably safeguarding source integrity for general quantum-information and communication tasks with Bayesian reasoning. Finally, we discuss how relative belief can be exploited to carry out parametric model certification and estimate the total dimension of the quantum state for the combined (measured) physical and interacting external systems described by the Tavis--Cummings~model.
	\end{abstract}
	
	\maketitle
	
	\section{Introduction}
	
	Quantum information processing requires sources of very high quality. Given a prepared quantum state~$\varrho$, for certain tasks, one may be contented with using scalable measurement copy numbers to estimate state properties with shadow tomography~\cite{Aaronson:2018shadow,Huang:2020predicting}, or perform state verification~\cite{Zhu:2023efficient,Han:2021optimal,Zhu:2019general,Zhu:2019efficientv,Flammia:2011direct} to certify the state fidelity. More generally, quantum tomography~\cite{Chuang:2000fk,lnp:2004uq,Teo:2011me,Teo:2015qs,Zhu:2014aa} is indispensable when a thorough knowledge about the prepared source is needed to scrutinize a quantum task. For a low-rank~$\varrho$, using fewer measurement outcomes that are still informationally complete for state reconstruction is possible with modern compressive tomography~\cite{Teo:2021modern,Gil-Lopez:2021universal,Teo:2021benchmarking,Teo:2020cs,Kim:2020aa,Gianani:2020aa,Ahn:2019aa,Ahn:2019ns} that do not rely on any assumptions or measurement restrictions apart from the system's dimension, unlike conventional compressed-sensing-based schemes~\cite{Steffens:2017cs,Baldwin:2016cs,Goyeneche:2015aa,Kalev:2015aa,Gross:2010cs}.
	
	As general quantum-information processing tasks operate on a fixed Hilbert-space dimension, it is therefore evident that state-reconstruction protocols should also be endowed with the capability to estimate the correct physical dimension of quantum systems. One important example of profound importance is the use of spatial-mode photonic sources in quantum cryptography~\cite{Zapatero:2023advances,Portmann:2022security,Pirandola:2020advances}, where the premise of an entangled photon source is almost always taken for granted by a typical state tomography protocol. Such a methodology is unsatisfactory in assuring an uncompromisable source integrity against~eavesdroppers. A more general and appealing desire might be to estimate the number of degrees of freedom of an external system interacting with the physical source of interest that results in decoherence~\cite{Schlosshauer:2019quantum,Heusler:2018modeling,Rauch:1999decoherence,Dragan:2005depolarization,Bayat:2006threshold,Karpinski:2008fiber,Amaral:2019characterization} and losses~\cite{Mogilevtsev:2010single-photon,Carpenter:2013nonlinear,Shomroni:2014all-optical,Bonneau:2015on-demand,Mendoza:2016active,Jones:2018PolarizationDL,Omkar:2020resource-efficient,Bartolucci:2021switch,Omkar:2022all-photonic,Kim:2022quantum}, \emph{without} actually measuring the external system, thereby facilitating the ``shift in the Heisenberg cut'', as the phrase goes~\cite{Englert:2013quantum,Englert:1999remarks}, by incorporating more external parameters that are uncovered by RB into the quantum-mechanical formalism if so desired.
	
	For these and other quantum-information tasks that require proper hypotheses evaluation, we introduce a purely evidence-based Bayesian paradigm of \emph{relative belief}~(RB)~\cite{Al-Labadi:2023how,Evans:2023resolving,Englert:2021checking,Nott:2021using,Nott:2020checking,Evans:2020measurement,Evans:2018measuring,Al-Labadi:2018statistical,Evans:2016aa,Evans:2015measuring} that draws conclusions about a set of hypotheses about a given quantum system directly from experimental data probing the~system. It starts by accepting a set of prior probabilities as belief measures for the individual hypotheses before the experiment and compares their corresponding posterior probabilities after incorporating the data. The \emph{plausible} hypotheses that are supported by the data are those with posterior probabilities larger than the prior ones. This general statistical framework applies to \emph{all} quantum systems, measurements and datasets of any sample size with no artificially-imposed assumptions or restrictions. To the authors' knowledge, the construction of plausible Bayesian error regions~\cite{Shang:2013cc,Li:2016da,Teo:2018aa,Oh:2018aa,Oh:2019efficient,Oh:2019probing,Sim:2019proper} in state and parameter estimation, and data analyses for finding evidence against local hidden-variable models~\cite{Gu:2019very} are the only avenues where RB was invoked in quantum mechanics and quantum information theory thus far. In view of this, the article aims to establish this paradigm as a key standard for assessing quantum systems.
	
	For hypotheses involving quantum models of different dimensions, the RB methodology functions as a model-selection~protocol. After an overview of the basic elements in Bayesian reasoning and a formal introduction to the RB framework in Sec.~\ref{sec:RBframework}, we shall first establish the RB protocol as the more conservative one in comparison to a broad class of information-criterion-based methods, including Akaike's information criterion~\cite{Burnham:2002model,Stoica:2004model-order}, in Sec.~\ref{sec:RBconserv}. More specifically, for sufficiently large datasets, we show that RB typically chooses a model dimension no smaller than those from any of the information-criterion-based methods. 
	
	Next, in Sec.~\ref{sec:imp_sources} we apply the RB methodology to track the quality of photon sources, which is a major state-preparation step in quantum-information and communication tasks. We demonstrate that with realistic lossy photon-number-resolving detectors of limited resolution, RB can successfully certify whether a specified photon source indeed produces a certain number of photons or is erroneously generating more photons than it should. Such a purely evidence-based reasoning presents a relatively more reliable strategy to test the integrity of photon~sources. 
	
	As a testament to the versatility of RB, in Sec.~\ref{sec:decoherence} we showcase its capability in estimating the degrees of freedom of an external system coupled to another physical system that an observer can access and measure. Clearly, such a precise assessment of the external system is only possible if relevant information about the number of interacting degrees of freedom is nontrivially imprinted on the reduced state of the measured physical system and the number of external degrees of freedom is manageable for their precise mathematical modeling based on a fair understanding of the physical interaction. We elaborate on two cases by analyzing the performance of RB with simplified parametric light-matter interaction models that are caricatures of typical physical situations in which the external system is, in reality, an environmental bath with an astronomical number of degrees of freedom. The interaction unitary operators for these simplified models take the variant forms of the extended Jaynes--Cummings model~\cite{Vivek:2023nonequilibrium,Cheng:2023quantum,Cui:2023effective,Nodurft:2019optical,Ezaki:1995photon,Seke:1985extended,Jaynes:1963comparison}. We show that in the case where the measured system is a two-level atom interacting with photon-field modes, RB is unable to discern the number of interacting field modes as these modes may be collectively describe by one single mode operator obeying the same bosonic commutation relations. However, when the measured system is a single-mode photon field interacting with multiple two-level molecular absorbers---the Tavis--Cummings model~\cite{Tavis:1968exact,Tavis:1969approximate}---then there exists nontrivial information about the number of absorbers encoded in the photonic reduced state that is accessible to RB through the photonic measurement data.
	
	\section{Relative belief in Bayesian reasoning}
	\label{sec:RBframework}
	
	\subsection{Basic framework}
	The concept of Bayesian statistical inference revolves around the mathematical encapsulation of beliefs for a set of hypotheses, which, in our case, concern a given quantum system of interest. For notational simplicity, we shall consider a discrete set of hypotheses, although the same treatment in this section may be applied to a continuous hypotheses set~(see Appendix~\ref{app:RBcont} for the more general expressions). Suppose one wishes to evaluate a set of $K$ hypotheses $\{\mathcal{H}_1,\mathcal{H}_2,\ldots,\mathcal{H}_K\}$ about the system. Before any experiment is performed on this system, one may harbor preconceived beliefs about each hypothesis. The initial strength of belief that $\mathcal{H}_k$ is correct is quantified by the \emph{prior probability} $\PR(k)$. After which, an experiment is conducted to acquire a measurement dataset $\mathbb{D}$ from the quantum system. The \emph{posterior probability} about $\mathcal{H}_k$, $\PR(k|\mathbb{D})$, then quantifies the strength of belief that this hypothesis is correct after reviewing the dataset $\mathbb{D}$. According to Bayes's theorem, we have the relation
	\begin{equation}
		\PR(k|\mathbb{D})=\dfrac{L(\mathbb{D}|k)\PR(k)}{\sum^K_{k'=1}L(\mathbb{D}|k')\PR(k')}
	\end{equation}
	between the belief strength before and after obtaining the dataset, where $L(\mathbb{D}|k')$ is the \emph{likelihood} of obtaining $\mathbb{D}$ given~$\mathcal{H}_k$.
	
	\emph{Relative belief} directly compares the prior and posterior belief strengths, giving us the RB ratio
	\begin{equation}
		\RB(k)=\dfrac{\PR(k|\mathbb{D})}{\PR(k)}=\dfrac{L(\mathbb{D}|k)}{\sum^K_{k'=1}L(\mathbb{D}|k')\,\PR(k')}\,,
		\label{eq:RBratio}
	\end{equation} 
	to ascertain whether the $k$th hypothesis, $\mathcal{H}_k$, is \emph{plausible}; that is, it is supported by $\mathbb{D}$~\cite{Evans:2015measuring} in accordance with an increased belief measure. If so, then $\RB(k)>1$, which states that the belief in favor of $\mathcal{H}_k$ is stronger after acquiring $\mathbb{D}$. Otherwise, one either has $\RB(k)<1$ or the inconclusive case where $\RB(k)=1$ precisely, the latter of which almost surely never occurs in typical situations as it would, otherwise, imply the mutual statistical independence of $\mathbb{D}$ and $\mathcal{H}_k$.
	
	The RB paradigm for deciding which hypothesis is supported by the data has a built-in consequence, namely that any hypothesis possessing the largest likelihood, say the $k_{\textsc{ml}}$th hypothesis with a likelihood value of $L_\mathrm{max}\equiv L(\mathbb{D}|k_{\textsc{ml}})=\max_{k'}L(\mathbb{D}|k')$, will always give $\RB(k_{\textsc{ml}})>1$:
	\begin{align}
		\RB(k_{\textsc{ml}})=&\,\dfrac{L_\mathrm{max}}{\sum^K_{k'=1}L(\mathbb{D}|k')\,\PR(k')}\nonumber\\
		>&\,\dfrac{L_\mathrm{max}}{L_\mathrm{max}\sum^K_{k'=1}\PR(k')}=1\,.
		\label{eq:RBmlgeq1}
	\end{align}
	The strict inequality holds so long as hypotheses of smaller likelihoods are included in the RB~test. Note that this property extends to the continuous case where $K=\infty$ and $\PR(k')$ are probability density functions. Hence, for a bounded set of hypotheses where all but the most-likely hypotheses fail the RB test, the most-likely hypotheses should not be immediately taken to be the valid ones for the quantum~system. Rather, the use of common sense is in order to decide \emph{whether or not} this observation is an indication that an extension of the current set of hypotheses is necessary. This ensures that the observer does not pick the wrong hypotheses from a highly-restricted set and incorrectly rule out untested~hypotheses. A simple example illustrating the exercise of this precaution is presented in Sec.~\ref{subsec:RBDC}.
	
	\subsection{Evidence strength}
	
	It is necessary to compare the \emph{evidence strengths} amongst all obtained RB ratios. A measure of how strong our conviction is that the evidence supports a given hypothesis is the conditional probability
	\begin{equation}
		E(k)=\sum_{k'\neq k}\PR(\RB(k')\leq\RB(k)|\mathbb{D})\,,
	\end{equation}
	that is, the total posterior probability of all other hypotheses giving an RB ratio no larger than $\RB(k)$~\cite{Evans:2015measuring}. 
	
	\begin{figure}[t]
		\includegraphics[width=0.85\columnwidth]{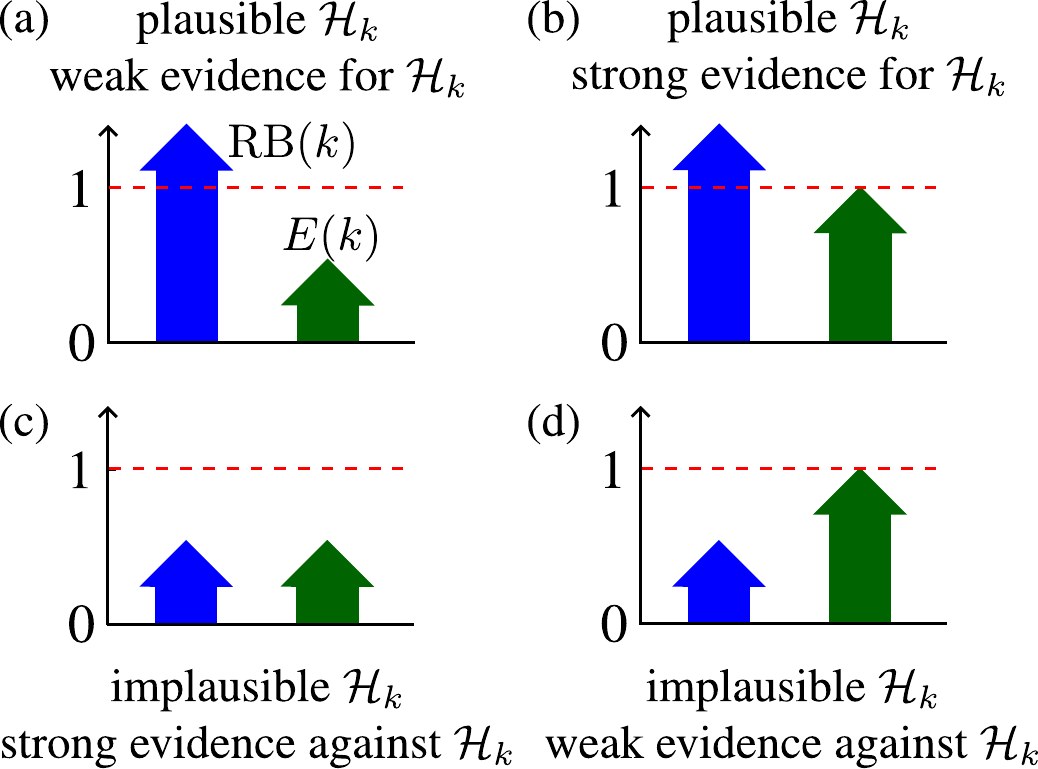}
		\caption{\label{fig:Ek}Interpretation of $E(k)$ depending on the magnitude of $\RB(k)$ for each hypothesis in the set $\{\mathcal{H}_1,\mathcal{H}_2,\ldots,\mathcal{H}_K\}$.}
	\end{figure}
	
	Depending on whether $\RB(k)$ is greater or less than 1, the interpretation of $E(k)$ is as given in Fig.~\ref{fig:Ek}. Thus, an \mbox{$\RB(k)>1$} accompanied by a large $E(k)$ signifies a stronger conviction that the true hypotheses lie in the set $\{k'|\RB(k')\leq\RB(k)\}$ and indicates a strong evidence in favor of the plausible $\mathcal{H}_k$, whereas both $\RB(k)<1$ and a small $E(k)$ represent a strong belief that the true hypotheses is in $\{k'|\RB(k')>\RB(k)\}$, and hence a strong evidence against the implausible~$\mathcal{H}_k$.
	
	\subsection{Plausible intervals and regions}
	
	If each hypothesis $\mathcal{H}_k$ is representable by a numerical value~$y_k$, then one can additionally construct geometrically-visualizable error intervals around each value. In Bayesian terms, these are commonly known as \emph{credible intervals} for single numerical values, or more generally \emph{credible regions} when referring to multidimensional parameters. Very generally, around $y_k$, the interval $\mathcal{R}_{\Delta_1,\Delta_2}=y_{k-\Delta_1}\leq y_k\leq y_{k+\Delta_2}$, defined by nonnegative integers~$\Delta_1$ and $\Delta_2$ for instance, of \emph{size} (prior content) $\mathcal{S}_{\Delta_1,\Delta_2}(y_k)=\sum^{k+\Delta_2}_{l=k-\Delta_1}\PR(y_{l})\leq1$ would have a \emph{credibility} (posterior content) defined by $\mathcal{C}_{\Delta_1,\Delta_2}(y_k)=\sum^{k+\Delta_2}_{l=k-\Delta_1}\PR(y_{l}|\mathbb{D})\leq1$, which is the probability that this interval contains the value $y_k$ given the dataset~$\mathbb{D}$.
	
	Being a mechanism for Bayesian reasoning, RB is also compatible with such intervals. However, the notion of evidence-based belief introduces another type of Bayesian interval that is more telling. For this purpose, the observer first orders the hypothesis set $\{y_1,y_2,\ldots,y_K\}$ in \emph{increasing} likelihood, namely $L(\mathbb{D}|y_k)\leq L(\mathbb{D}|y_{k'})$ whenever $k<k'$. After such an ordering, it is clear, from~\eqref{eq:RBratio}, that \mbox{$\RB(y_k)\leq \RB(y_{k'})$}. Next, the observer is ready to construct \emph{plausible intervals}, which are those containing \emph{only} the plausible hypotheses. If $k_\mathrm{eff}=\arg\min_{k}\{\RB(y_k)\,\text{or}\,L(\mathbb{D}|y_k)\,|\,\RB(y_k)>1\}$, then for some nonnegative integer $\Delta$,
	\begin{align}
		\mathcal{R}_{\mathrm{plaus},\Delta}=&\,\{y_{k_\mathrm{eff}},y_{k_\mathrm{eff}+1},\ldots,y_{k_\mathrm{eff}+\Delta}\}\,,\nonumber\\
		\mathcal{S}_{\mathrm{plaus},\Delta}=&\,\sum^{k_\mathrm{eff}+\Delta}_{k=k_\mathrm{eff}}\PR(y_{k})\,,\nonumber\\
		\mathcal{C}_{\mathrm{plaus},\Delta}=&\,\sum^{k_\mathrm{eff}+\Delta}_{k=k_\mathrm{eff}}\PR(y_{k}|\mathbb{D})>\mathcal{S}_{\mathrm{plaus},\Delta}\,.
	\end{align}
	The largest plausible interval with $\Delta=K$ corresponds to those developed in~\cite{Shang:2013cc,Li:2016da,Teo:2018aa,Oh:2018aa,Oh:2019efficient,Oh:2019probing,Sim:2019proper}. It is clear that any credible interval larger than $\mathcal{R}_\mathrm{plaus}$ would necessarily contain implausible hypotheses that are unsupported by the data.
	
	\subsection{Error probabilities}
	
	We may also analyze the error probabilities of wrongfully accepting $\mathcal{H}_k$ to be plausible when it is not an element of the set of true hypotheses, $\mathcal{T}$. This \emph{type-I} probability is given by the sum of \emph{prior probabilities} under this situation:
	\begin{equation}
		\epsilon_\mathrm{I}=\sum_{\mathcal{H}_k\notin\mathcal{T}}\PR(k)\,\eta(\RB(k)-1)\,,
		\label{eq:typeI}
	\end{equation}
	where $\eta(\,\,\bm{\cdot}\,\,)$ is the Heaviside step function. By definition, a ``true hypothesis'' is one that maximizes the likelihood in the infinite data-sample limit. In quantum tomography, for instance, the ``true state'' is the maximum-likelihood~(ML) reconstructed state~\cite{Banaszek:1999ml,Fiurasek:2001mq,lnp:2004uq,Rehacek:2007ml,Teo:2011me,Teo:2015qs} in this asymptotic limit. This implicitly assumes statistical consistency in hypothesis (parameter) estimation and that the underlying POVM and hypothesis models have passed auxiliary procedures such as model checking and prior-data-conflict analyses~\cite{Englert:2021checking,Nott:2020checking,Al-Labadi:2018statistical}, the elaboration of which is beyond the scope of this article.
	
	The opposite situation, in which $\mathcal{H}_k$ is wrongfully deemed implausible when it is actually true, results in the \emph{type-II} error~probability
	\begin{equation}
		\epsilon_\mathrm{II}=\sum_{\mathcal{H}_k\in\mathcal{T}}\PR(k)\,\eta(1-\RB(k))\,.
		\label{eq:typeII}
	\end{equation}
	These two error probabilities measure the reliability of the RB~strategy.
	
	\begin{figure*}[t]
		\includegraphics[width=1.5\columnwidth]{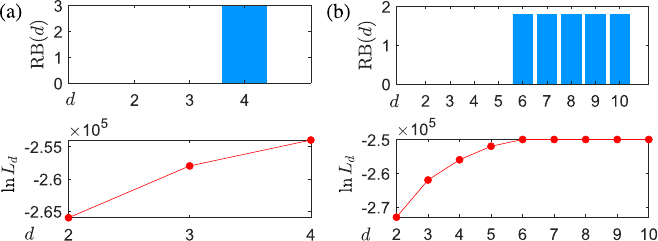}
		\caption{\label{fig:toy}Plots of RB ratios and log-likelihood values for Hilbert spaces of various~$d$ describing the unknown ten-dimensional $\varrho\,\widehat{=}\,\mathrm{diag}(1,1,1,1,1,1,0,0,0,0)/6$. The dimension sets (a)~$\{d_1=2,d_2=3,d_3=4\}$ and (b)~$\{d_1=2,d_2=3,d_3=4,d_4=5,d_5=6,d_6=7,d_7=8,d_8=9,d_9=10\}$ are tested. The POVM is a fixed set of randomly-chosen 11~von~Neumann bases used to gather a multinomial dataset $\mathbb{D}$ with a total of $N=10^4$ copies.}
	\end{figure*}
	
	Both $\epsilon_\mathrm{I}$ and $\epsilon_\mathrm{II}$ depend on the number of sampling copies~$N$ used to collect the measurement dataset~$\mathbb{D}$. In the large-$N$ limit, the posterior probabilities $\PR(k|\mathbb{D})$ peaks very sharply at only the true hypotheses, so that the step function in Eq.~\eqref{eq:typeI} is~1 almost only for $\mathcal{H}_k\in\mathcal{T}$. Put differently, the step function in Eq.~\eqref{eq:typeII} is almost always zero for $\mathcal{H}_k\in\mathcal{T}$ in this limit. Therefore, in the limit of large sampling copies, both error probabilities must converge to~zero.
	
	\subsection{Dimension certification}
	\label{subsec:RBDC}
	
	While RB is a general statistical paradigm that applies to any probabilistic scenario, to put things into a context that shall be relevant to subsequent discussions and is also considered in the companion Letter~\cite{Teo:2024evidence-based}, we shall explicitly discuss the \emph{relative-belief dimension certification} (RBDC) protocol that caters to the certification of model dimensions for describing the state of the quantum system. The set of hypotheses of interest is then the \emph{ordered} set of $K$ dimensions of Hilbert spaces ($d_1<d_2<\ldots<d_K$) describing the quantum state~$\varrho$, then $\PR(d_k)\equiv\PR(k)$ is the prior probability for the Hilbert space of dimension~$d_k$. In this context, the likelihood $L_{d_k}\equiv L(\mathbb{D}|k)=\max_{\varrho(d_k)} L(\mathbb{D}|\varrho(d_k))$ consequently takes the ML value over all $d_k$-dimensional quantum states for the dataset~$\mathbb{D}$.
	
	A key feature with regard to both the likelihood and RB ratio in dimension certification is that if, for some $k=k_0$, \mbox{$\RB(d_{k_0})>1$}, then all dimensions $d>d_{k_0}$ must also result in $\RB(d>d_{k_0})>1$. The reason is because if an ML state estimator $\widehat{\varrho}_\textsc{ml}(d_{k'})$ is found for \emph{any} $d_{k'}$-dimensional Hilbert space, then any other Hilbert space larger than this one must, obviously, contain $\widehat{\varrho}_\textsc{ml}(d_{k'})$, so that $L_{d_{k>k'}}\geq L_{d_{k'}}$. Therefore, both the likelihood and RB ratio never decreases with increasing model dimension.
	
	The procedure of RBDC is as follows:
	\begin{enumerate}
		\item Choose a finite set of Hilbert-space dimensions $\{d_1,d_2,\ldots,d_{K}\}$ to be tested with RB.
		\item Assign a prior probability $\PR(d_k)$ to each dimension.
		\item Perform a measurement on $\varrho$ using a positive operator-valued measure~(POVM) and obtain a dataset $\mathbb{D}$.
		\item Choose a basis representation that typically depends on the physical task at hand (for photonic systems, it could either be the energy basis, angular-momentum basis, \emph{et~cetera}).
		\item Starting with $k=1$:
		\begin{itemize}
			\item[(i)] Truncate the POVM outcomes into $d_k\times d_k$ matrices.
			\item[(ii)] Calculate $L_{d_k}=\max_{\varrho(d_k)}\{L(\mathbb{D}|\varrho(d_k))\}$ and store this value. This value will typically be minuscule, so a high-precision storage format~\cite{DErrico:2018hpf} will be needed.
			\item[(iii)] Increase $k$ by 1.
			\item[(iv)] Repeat (i)--(iii) for all $1\leq k\leq K$.
		\end{itemize}
		\item Compute and store the posterior probabilities $\PR(d_k|\mathbb{D})$.
		\item Compute $\RB(d_k)$ for all $1\leq k\leq K$ and find $\dRB$, the smallest plausible dimension. All larger dimensions will also be plausible.
		\item Compute $E(d_k)$ for all $1\leq k\leq K$, and $\mathcal{C}_{\mathrm{plaus},\Delta}$.
	\end{enumerate}
	
	From hereon, we suppose that the POVM is the set of positive outcome operators $\{\Pi_j\}^M_{j=1}$ with the number of outcomes ($M$)~fixed. This set may constitute multiple von~Neumann bases that are weighted equally, such that the sum \mbox{$\sum^M_{j=1}\Pi_j=1$} always holds. We emphasize here that even a single von~Neumann basis is sufficient for ascertaining the~$\deff$ of~$\varrho$ from its diagonal-element estimation, as the positivity of $\varrho$ requires all off-diagonal terms corresponding to null diagonal elements also be~zero. In subsequent discussions, all simulation data click frequencies are generated under the assumption that each outcome is detected independently, so that the click frequencies $n_j$, such that $\sum^M_{j=1}n_j=N$, of a fixed sampling-copy number~(sample size)~$N$ follow the multinomial distribution with measurement probabilities $p_{d,j}=\tr{\varrho(d)\,\Pi_j}$ for all $1\leq j\leq M$ and a fixed truncated-Hilbert-space dimension~$d$. It is important to note that the space for which $\sum^M_{j=1}p_{d,j}=1$ is \emph{not} the entire probability simplex, but a (typically) much smaller subspace governed by $\varrho(d)\in\mathcal{Q}_d$ that are positive and unit trace, where $\mathcal{Q}_d$ is the $d^2$-dimensional quantum state~space. In the limit of large sampling-copy number (sample-size)~$N\rightarrow\infty$, we have \mbox{$n_j\rightarrow N p_{d,j}$}. The corresponding likelihood for a \emph{particular} detected POVM-outcome sequence is \mbox{$L(\mathbb{D}|\varrho(d))=\prod^M_{j=1}p^{n_j}_{d,j}$}.
	
	The value $L_{d_k}$, in statistics, is also known as the profile likelihood as it is the likelihood of a specific state~(the ML state) in the state space. In common experimental situations, where data samples are typically large, the RBDC scheme involving~$L_{d_k}$ that is established in this section is asymptotically identical to a slightly more general methodology~(presented in Appendix~\ref{app:fullRBDC}), the latter of which is also applicable to very small data samples but requires much more sophisticated state-space sampling~\cite{Shang:2015mc,Seah:2015mc} procedures than just likelihood~maximization. The simplicity in maximizing the likelihood over state-space averaging is an attractive feature that makes this RBDC scheme more readily operational.
	
	We end this section by commenting more concretely on the implication of~\eqref{eq:RBmlgeq1} in dimension certification. Consider a ten-dimensional quantum state $\varrho\,\widehat{=}\,\mathrm{diag}(1,1,1,1,1,1,0,0,0,0)/6$ is sparse in the computational basis and can effectively be contained in a six-dimensional Hilbert space, whereas only the dimension set $\{d_1=2,d_2=3,d_3=4\}$ is tested with RBDC. Then for a sufficiently large dataset such that statistical fluctuation is low, Fig.~\ref{fig:toy}(a) shows that $\RB(d_1)=0=\RB(d_2)$ and $\RB(d_3)>1$ with a dataset from a randomly-chosen POVM. The RB ratios for $d_1$ and $d_2$ are (almost) identically zero since the respective Hilbert spaces are too small to contain~$\varrho$ in the computational basis. That the last RB ratio is greater than one is the result of $\log L_{d_3}=\max_{1\leq k\leq 3}\{\log L_{d_k}\}^3_{k=1}$ as in Fig.~\ref{fig:toy}(a), and is, thus, consistent with~\eqref{eq:RBmlgeq1}. Naively, we may take $\deff=d_3=4$ at face value and conclude that $\deff=4$ is the smallest Hilbert space that can describe~$\varrho$. However in this case, it is clear that this artifactual conclusion is false as $\varrho$ cannot be adequately contained in a four-dimensional Hilbert space also. Rather, Fig.~\ref{fig:toy}(a) is an indication that further certification with a larger dimension set is necessary. Indeed, extending the set to $d=10$ leads to the right assessment, as demonstrated in Fig.~\ref{fig:toy}(b).
	
	
	\section{Conservativeness of relative-belief dimension certification}
	\label{sec:RBconserv}
	
	For a given number $N$ of sampling copies used to obtain the dataset $\mathbb{D}$, there exist other well-known model-dimension selection methods that minimize a class of information functions in the dimension~$d$ of the form $I_{\alpha,d}=\alpha\kappa_d-\log L_d$ with a positive~$\alpha$ that scales sublinearly with~$N$. Here, $\kappa_d$ is a function that is \emph{strictly increasing} in~$d$. When applied to quantum-state tomography, for instance, $\kappa_d=d^2-1$. Special cases of this class includes Akaike's~($\alpha=1$, AIC) and the Bayesian~[$\alpha=(\log N)/2$, BIC] information criteria~\cite{Burnham:2002model,Stoica:2004model-order}. The effective dimension $d_I$ refers to the dimension corresponding to the global minimum of $I_{\alpha,d}$. If there exist multiple dimensions giving this global minimum value, then we pick the effective dimension~$d_I$ to be the largest global-minimum one as the conservative choice for such an information criterion. 
	
	It turns out that in the large-$N$ limit, even with such a choice, if~$\PR(d)>0$, then RBDC typically announces a~\mbox{$\dRB\geq d_I$} with very high probability, thereby rendering RBDC as the more conservative quantum protocol for ascertaining the effective dimension of an unknown state in contrast to this class of information-criterion-based strategies. 
	
	The remaining part of this section provides an argument for this fact. To this end, we stress that whether the global minimum of $I_{\alpha,d}$ is unique or not, as $d_I$ is the largest dimension for which $I_{\alpha,d}$ is minimized, the strict inequality,
	\begin{equation}
		\alpha\,\kappa_{d_I}-\log L_{d_I}<\alpha\kappa_{d_I+1}-\log L_{d_I+1}\,,
	\end{equation}
	holds.
	
	For a set of $K$ dimensions $\{d_1,d_2,\ldots,d_K\}$, if $L_\mathrm{max}=\max_{d_k}\{L_{d_k}\}$, then $\log L_{d_I}\leq\log L_\mathrm{max}$, which~gives
	\begin{equation}
		\alpha\,\kappa_{d_I}-\log L_\mathrm{max}<\alpha\kappa_{d_I+1}-\log L_{d_I+1}\,,
	\end{equation}
	or
	\begin{equation}
		\log L_{d_I+1}<\log L_\mathrm{max}+\alpha(\kappa_{d_I+1}-\kappa_{d_I})\,.
		\label{eq:LdI+1_Lmax}
	\end{equation}
	
	On the other hand, an~$\RB(d=\dRB)>1$ implies that
	\begin{equation}
		\log L_{\dRB}>\log\left(\sum^{K}_{k=1}L_{d_k}\PR(d_k)\right)>\log(L_{d_0})+\log(\PR(d_0))\,,
	\end{equation}
	where~$d_0=\arg\max_{d_k}\{L_{d_k}\,\PR(d_k)\}$ such that $\PR(d_k)\leq\PR(d_0)$ for all $1\leq k\leq K$. To complete the argument, we first assume, for simplicity, a uniform~$\PR(d_k)=1/K$, leading~to
	\begin{equation}
		\log L_{\dRB}>\log L_\mathrm{max}-\log K\,,
		\label{eq:LRB_Lmax}
	\end{equation}
	so that \eqref{eq:LdI+1_Lmax} and \eqref{eq:LRB_Lmax} bring us to the inequality chain
	\begin{equation}
		\log L_{d_I}\leq\log L_{d_I+1}<\log L_{\dRB}+\alpha(\kappa_{d_I+1}-\kappa_{d_I})+\log K\,.
		\label{eq:ineq}
	\end{equation}
	Since $\alpha$ is sublinear in~$N$ and both $\kappa_d$ and $K$ are independent of $N$, we find that when $N\gg1$, these terms are dominated by the log-likelihoods, which are of order $N$, such that $\log L_{d_I}<\log L_{\dRB}$ with a very large probability. Finally, the nondecreasing property of the likelihood in~$d$ tells us that $d_I<\dRB$. 
	
	Note that for astronomical datasets~($N\rightarrow\infty$), the likelihood function $L_d$ peaks extremely sharply at some saturated maximum value $L_{d_\mathrm{c}}$ that is achieved for $d\geq d_\mathrm{c}$, where $d_\mathrm{c}$ is some critical dimension; that is, $\log L_d$ drops drastically for $d<d_\mathrm{c}$. In this asymptotic case, we almost surely have $\log L_{d_I}=\log L_{d_\mathrm{c}}=\log L_{\dRB}$. The existence of such a saturation may arise either when $\varrho$ has a photon-number distribution that ends abruptly after certain photon number $d_\mathrm{c}-1$ (such as the situation for perfect Fock states and their superposition or mixtures, which is an idealized situation), or when the distribution tail becomes numerically indistinguishable from zero (that of a limited-intensity coherent state, or practically any physically-realizable quantum state for that matter). Then in this case, $d_I$ must be the smallest dimension for which $\log L_{d_I}=\log L_{d_\mathrm{c}}$, since $\alpha\kappa_{d_I}-\log L_{d_\mathrm{c}}<\alpha\kappa_{d_I+1}-\log L_{d_\mathrm{c}}$, which is obvious from the strictly-increasing $\kappa_d$. On the other hand, by definition of RBDC, we must also have $\dRB$ to be the smallest dimension for which $\log L_{\dRB}=\log L_{d_\mathrm{c}}$, which concludes that $d_\mathrm{I}=\dRB$ for such a situation.
	
	\begin{figure}[t]
		\includegraphics[width=1\columnwidth]{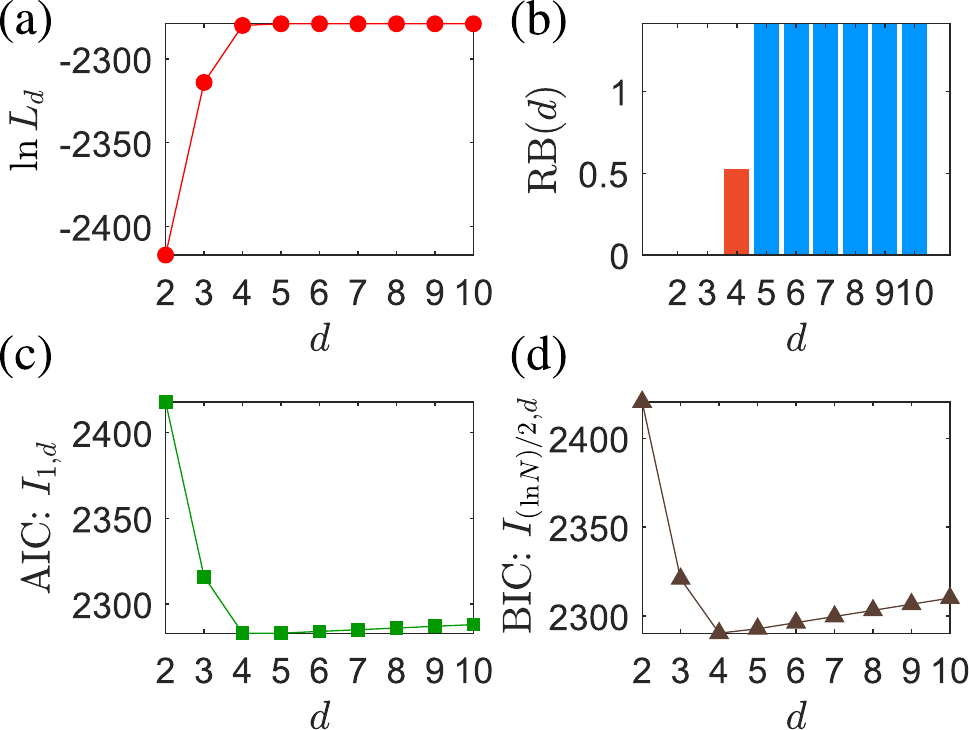}
		\caption{\label{fig:RBDC_AIC_BIC}(a)~Log-likelihood values for various~$d$ that lead to the performances of uniform-prior-based RBDC (red and blue bars refer to RB ratios smaller and greater than one), (b)~AIC and (c)~BIC based on a single-von~Neumann-basis dataset from measuring $N=1000$~copies of the unknown ten-dimensional $\varrho\,\widehat{=}\,\mathrm{diag}(1,1,1,1,1,1,0,0,0,0)/6$. The respective effective dimensions selected by each of these three methods may be readily read off as $\dRB=5$ and $d_{\mathrm{AIC}}=4=d_{\mathrm{BIC}}$ from the~graphs. The red RB-ratio bar also indicates that evidence is found \emph{against} the smaller $d_I=d_\mathrm{AIC}$ and $d_I=d_\mathrm{BIC}$~values.}
	\end{figure}
	
	For the case with arbitrary priors where $\PR(d_k)>0$ for any~$d_k$, it is easy to see that for~$N\gg1$, the variation of $L_{d_k}\,\PR(d_k)$ in $d_k$ is approximately that of $L_{d_k}$~[$\log L_{d_k}+\log \PR(d_k)\cong\log L_{d_k}$]. This is the \emph{data-dominant} situation in which the influence of the prior distribution is small compared to that of the data, with $N=\infty$ defining the extreme case of this situation. Under this situation, the above argument~follows.
	
	Figure~\ref{fig:RBDC_AIC_BIC} illustrates the performances of RBDC, AIC and BIC for a single dataset derived from one simulated measurement trial with $N=1000$ copies of $\varrho\,\widehat{=}\,\mathrm{diag}(1,1,1,1,1,1,0,0,0,0)/6$ using a single randomly-generated von~Neumann basis. The uniform prior distribution $\PR(d_k)=1/9$ for $2\leq d_k\leq 10$ is considered. Here, $\kappa_d=d-1$ since only $d-1$ independent diagonal elements are estimable upon Hilbert-space~truncation with such a measurement in this example.
	
	\begin{figure}[t]
		\includegraphics[width=0.47\columnwidth]{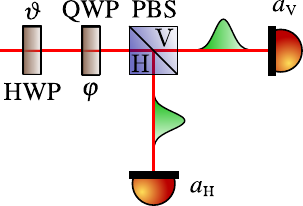}
		\caption{\label{fig:single-mode}Polarization measurement of multiphoton pulses for a single spatial mode using wave plates and PNRDs, where the mode operators on each arm are respectively $a_\textsc{h}$ and $a_\textsc{v}$.}
	\end{figure}
	
	\section{Relative-belief certification of imperfect sources}
	\label{sec:imp_sources}
	
	Quantum-state tomography on imperfect photonic sources is one important application where RBDC can be employed to assess the quality of these photon sources to be utilized for quantum-information and communication tasks. Tasks requiring sources emitting a specified number of polarization photons will underperform in quality in terms of state or computation output if the sources generate different numbers of photons. This has especially negative impact in quantum key distribution protocols when spurious additional polarization photons supplied by the source, unbeknownst to the user, could lead to security attacks by an eavesdropper~\cite{Scarani:2009security,Felix:2001faint}.
	
	In this section, we shall consider sources emitting photons in a certain number of spatial modes. Figure~\ref{fig:single-mode} shows the standard measurement setup for one spatial mode. It generally comprises a polarizing beam splitter~(PBS) and two wave plates HWP~($\vartheta$) and QWP~($\varphi$). At the end of the PBS, two photon-number-resolving detectors~(PNRDs) are present to measure multiphoton pulses at each~port. 
	
	The mathematical description of the operator corresponding to the measurement of $n$~photons on one of the output arms of the combined PBS and PNRDs of detection efficiency~$\eta\leq1$ and a maximum photon-number resolution of~$n_0$, excluding the wave plates and presence of dark counts, is given by the normal-order form~\cite{Sperling:2012true,Prasannan:2022direct} 
	\begin{align}
		\Pi^{(\eta)}_{n}=&\,\binom{n_0}{n}\bm{:}\left(1-\E{-\eta a^\dag a/n_0}\right)^n\E{-\eta(n_0-n) a^\dag a/n_0}\bm{:}\nonumber\\
		=&\,\binom{n_0}{n}\sum^n_{k=0}\binom{n}{k}(-1)^k\left[1-\eta\left(1+\frac{k-n}{n_0}\right)\right]^{a^\dag a}\,,
	\end{align}
	where the second equality is obtained using the identity \mbox{$\bm{:}\E{-\lambda a^\dag a}\bm{:}\,\,=(1-\lambda)^{a^\dag a}$}. These operators equivalently represent $n_0$ ``on-off'' detections, where the number of ``on'' detections is $n$. To understand the form of these operators, we consider the perfect situation of $\eta=1$. Then, $n_0=1$ gives rise to the familiar special case of a single ``on-off'' detection:
	\begin{align}
		\Pi^{(\eta=1)}_0=&\,\bm{:}\E{-a^\dag a}\bm{:}\,=0^{\,a^\dag a}=\vacket \vacbra \,,\nonumber\\
		\Pi^{(\eta=1)}_1=&\,1-\vacket \vacbra =\ket{1}\bra{1}+\ket{2}\bra{2}+\ldots\,.
	\end{align}
	When $n_0=2$, we have
	\begin{align}
		\Pi^{(\eta=1)}_0=&\,\vacket \vacbra \,,\nonumber\\
		\Pi^{(\eta=1)}_1=&\,2\left(\frac{1}{2^{a^\dag a}}-\vacket \vacbra \right)\,,\nonumber\\
		\Pi^{(\eta=1)}_2=&\,1-\frac{2}{2^{a^\dag a}}+\vacket \vacbra \,.
	\end{align}
	
	Operators governing the HWP and QWP wave plates are respectively expressed as the unitary operators $\E{-\I\varphi J_3}$ and $\E{-\I\vartheta J_2}$ in terms of the angular-momentum operators $J_1=(a_\textsc{h}^\dag a_\textsc{v}+a_\textsc{v}^\dag a_\textsc{h})/2$, $J_2=(a_\textsc{h}^\dag a_\textsc{v}-a_\textsc{v}^\dag a_\textsc{h})/(2\I)$ and $J_3=(a_\textsc{h}^\dag a_\textsc{h}-a_\textsc{v}^\dag a_\textsc{v})/2$, which form the Jordan--Schwinger bosonic representation of the SU(2) group generators~\cite{Goldberg:2021quantum,Schwinger:1952report,Jordan:1935Zusammenhang}. Upon defining $J_\pm=J_1\pm J_2$, the combined unitary operator of the HWP--QWP sequence \emph{on an incoming signal state} is $U_{\vartheta,\varphi}=\E{-\I\varphi J_3}\,\E{-\I\vartheta J_2}=\E{\frac{\vartheta}{2}\left(J_+\E{-\I\varphi}-J_-\E{\I\varphi}\right)}$ and the polarimetric POVM outcomes for a single spatial mode are
	\begin{equation}
		\Pi^{(\eta)}_{n_\textsc{h},n_\textsc{v}}(\vartheta,\varphi)=U_{\vartheta,\varphi}^\dag\Pi^{(\eta)}_{n_\textsc{h}}\otimes\Pi^{(\eta)}_{n_\textsc{v}} U_{\vartheta,\varphi}\,.
		\label{eq:QPol_POVM}
	\end{equation}
	
	In Appendix~\ref{app:linindep}, we shall show that the number of linearly-independent POVM outcomes stated in \eqref{eq:QPol_POVM} is \emph{at~most}
	\begin{align}
		M_\mathrm{pol}=&\,2\sum^{n_0-1}_{k=0}(k+1)^2+(n_0+1)^2\nonumber\\
		=&\,\dfrac{1}{3}(n_0+1)(2n_0^2+4n_0+3)\,,
		\label{eq:Mpol}
	\end{align}
	which is significantly smaller than the operator dimension $(n_0+1)^4$ of any state $\varrho$ in the $(n_0+1)^2$-dimensional Hilbert space, as such a POVM only probes the polarization sector of~$\varrho$ as every measurement outcome gives nonzero transition amplitudes between two-mode Fock states of the \emph{same} total photon numbers.
	
	\subsection{Single spatial-mode photon source}
	
	The single spatial-mode source considered here is a heralded one~\cite{Meyer-Scott:2020review} from the output of a type-II spontaneous parametric down-conversion~(SPDC), in which the corresponding (simplified) interaction Hamilton operator
	\begin{align}
		H_\mathrm{int}=&\,\,g(\E{\I\phi}L_++\E{-\I\phi}L_-)\,,\nonumber\\
		L_+=&\,\,a^\dag_\textsc{h}b^\dag_\textsc{v}-a^\dag_\textsc{v}b^\dag_\textsc{h}=L^\dag_-\,,
	\end{align}
	with coupling strength~$g$ and phase~$\phi$~\cite{Simon:2000optimal,Kok:2000postselected}. This leads to the relevant interaction unitary operator $U_\mathrm{int}=\E{-\I H t}=\E{-\I gt(\E{\I\phi}L_++\E{-\I\phi}L_-)}$ that can be disentangled into
	\begin{align}
		U_\mathrm{int}=&\,\E{-\I\E{\I\phi}\tanh(gt)\,L_+}\,\E{-2\log(\cosh(gt))\,L_0}\nonumber\\
		&\,\times\E{-\I\E{-\I\phi}\tanh(gt)\,L_-}
	\end{align}
	since the $\mathfrak{su}(1,1)$-algebraic commutation relations
	\begin{align}
		[L_-,L_+]=&\,2L_0\equiv a^\dag_\textsc{h}a_\textsc{h}+a^\dag_\textsc{v}a_\textsc{v}+b^\dag_\textsc{h}b_\textsc{h}+b^\dag_\textsc{v}b_\textsc{v}+2\,,\nonumber\\
		[L_0,L_\pm]=&\,\pm L_\pm
	\end{align}
	hold for the generators $L_\pm$ and $L_0$~\cite{Truax:1988bch}.
	
	\begin{figure}[t]
		\includegraphics[width=1\columnwidth]{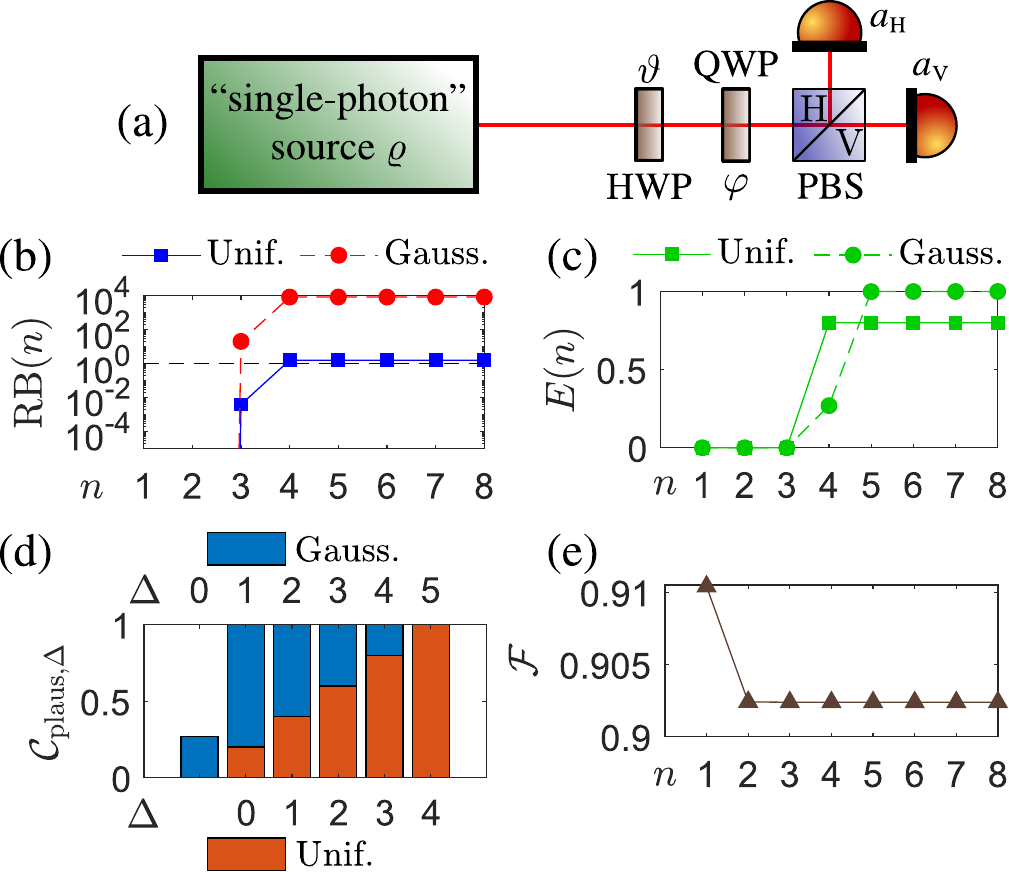}
		\caption{\label{fig:single-mode-source}Quality assessment and polarization-sector reconstruction of (a)~a supposed ``single-photon'' source in spatial mode~$a$ obtained from heralding with the lossy $\textsc{D}$ outcome in spatial mode~$b$ using RBDC. The (b)~RB ratios in vertically logarithmic scale, (c)~evidence strengths~$E(n)$ and (d)~credibilities of plausible intervals~$\mathcal{C}_{\mathrm{plaus},\Delta}$ for both the uniform and Gaussian~[$\propto \exp(-(n-1)^2)$] prior distributions are~plotted. (e)~Polarization-sector fidelities with $\mathcal{F}$ with $\ket{\text{1 ph}}_-$ are also shown. For each of the 20 randomly-chosen pairs of $0\leq\vartheta\leq\pi$ and $0\leq\phi\leq 2\pi$, a total of $N=10^4$ sampling copies are used to collect the measurement data. The label $n$ refers to the number of photons describing each polarization ket in Fock representation. The corresponding Hilbert-space model dimension is $d_n=(n+1)^2$.}
	\end{figure}
	
	The action of $U_\mathrm{int}$ on the vacuum state~$\vacket \vacbra $, thus, produces the pure quantum state~$\varrho_\textsc{SPDC}=\ket{\text{TMSV}}\bra{\text{TMSV}}$ described by the two-mode squeezed vacuum~(TMSV) ket
	\begin{equation}
		\ket{\text{TMSV}}=\sum^\infty_{n=0}\sum^n_{m=0}\ket{n-m,m,m,n-m}(-1)^m\frac{(\tanh \tau)^n}{(\cosh\tau)^2}
	\end{equation}
	in terms of two spatial modes of polarization degrees of freedom~\cite{Lamas-Linares:2001stimulated}, where the squeezing strength~$\tau=|gt|$ and the labels in the ket argument are ordered as $a_\textsc{h}$, $a_\textsc{v}$, $b_\textsc{h}$ and $b_\textsc{v}$ in optical modes. With perfect lossless PNRDs, the single spatial-mode heralded source may be produced by heralding one of the two linearly-polarized measurement outcomes in mode~$b$, that is, either the outcome ket $\ket{\textsc{D}}_b=(b^\dag_\textsc{h}+b^\dag_\textsc{v})\vacket /\sqrt{2}$ or $\ket{\textsc{A}}_b=(b^\dag_\textsc{h}-b^\dag_\textsc{v})\vacket /\sqrt{2}$, to obtain the respective perfect single-photon kets $\ket{\text{1 ph}}_\mp=(\ket{1}_\textsc{h}\ket{0}_\textsc{v}\mp\ket{0}_\textsc{h}\ket{1}_\textsc{v})/\sqrt{2}$ in mode~$a$. We choose $\tau=\log(2+\sqrt{3})/2\approx0.658\equiv 5.71~\mathrm{dB}$ that would give the maximum single-photon heralding success rate of~$\left[\tanh \tau/(\cosh\tau)^2\right]^2\approx14.8\%$.
	
	For a more realistic simulation, we shall, however, consider lossy detectors of efficiency $\eta=0.9$, so that the Fock state for every polarization mode,
	\begin{equation}
		\ket{n}\bra{n}\mapsto\widetilde{\ket{n}\bra{n}}=\eta^n\sum^\infty_{l=n}\ket{l}\binom{l}{n}(1-\eta)^{l-n}\bra{l}\,,
	\end{equation}
	becomes mixed under the photonic amplitude-damping (or photon-loss) channel~\cite{Albert:2018performance,Bergmann:2016quantum,Kiss:1995compensation,Lee:1993external}. The resulting imperfectly-heralded single spatial-mode output state in mode~$a$, $\varrho=\ptr{b}{\widetilde{\ket{\textsc{D}}_b{\vphantom{\ket{\textsc{D}}}}_b\!\bra{\textsc{D}}}\varrho_{\textsc{SPDC}}}$, will then contain contributions from higher photon numbers.
	
	For PNRDs with a maximum photon-number resolution of $n_0=8$, a total of 20 randomly-chosen wave-plate angle pairs $\vartheta,\varphi$ is found to give the maximum number of linearly independent measurement outcomes, namely $L_\mathrm{pol}=489$. We remark here that the number of sampling copies, $N$, is the number of pump pulses used to produce the SPDC source, so that a proper calibration may also allow us to measure the ``no~click'' counts~\cite{Nunn:2021heralding}. Therefore, we have a set of POVM that can fully characterize the polarization sector of~$\varrho$. 
	
	Figure~\ref{fig:single-mode-source} shows that the imperfectly-heralded state~$\varrho$ can be adequately described by polarization Fock kets of effective dimensions $\dRB=4$~($n_\mathrm{RB}=3$) and $\dRB=5$~($n_\mathrm{RB}=4$) for the Gaussian and uniform priors respectively, telling us that the source is emitting up to 4~photons with very high evidence strength based on the data and, thus, indicating the integrity deviation of the heralded source away from a perfect single-photon quantum~system. 
	
	\subsection{Double spatial-mode photon source}
	
	\begin{figure}[t]
		\includegraphics[width=1\columnwidth]{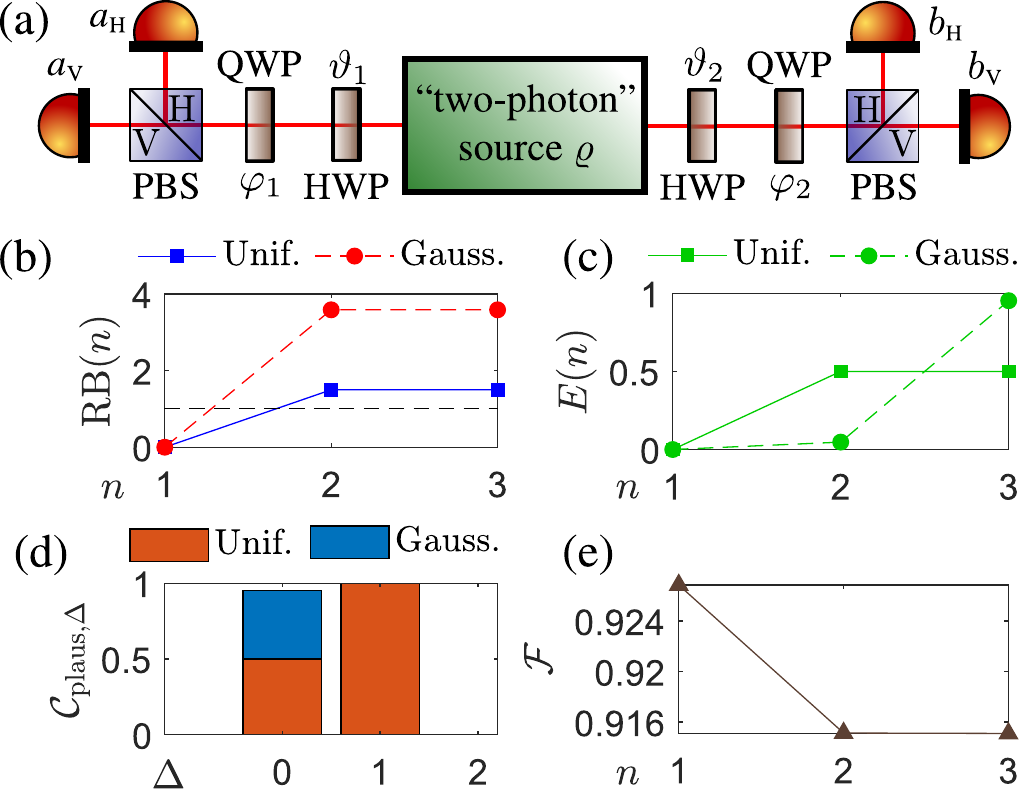}
		\caption{\label{fig:two-mode-source}Quality assessment and polarization-sector reconstruction of (a)~a supposed ``two-photon'' source emitting photons in two spatial modes (described by mode operator pairs $a_\textsc{h},a_\textsc{v}$ and $b_\textsc{h},b_\textsc{v}$ respectively) with RBDC, where (b)~the RB ratios, (c)~evidence strengths~$E(n)$ and (d)~credibilities of plausible intervals~$\mathcal{C}_{\mathrm{plaus},\Delta}$ for both the uniform and Gaussian~[$\propto \exp(-(n-1)^2)$] prior distributions are~plotted. (e)~The source quality is reflected by the polarization-sector fidelities $\mathcal{F}$ with the ideal ket $\ket{\text{2 ph}}=(\ket{1,0,0,1}-\ket{0,1,1,0})/\sqrt{2}$, where the purely-vacuum component in all ML state estimators is removed (followed by trace renormlization) to compute $\mathcal{F}$. For each measurement setting $(\vartheta_1,\varphi_1,\vartheta_2,\varphi_2)$, a total of $N=10^4$ sampling copies are used to collect the measurement data. Here, $n$ refers to the number of photons describing each polarization ket of every spatial mode in Fock representation. The corresponding Hilbert-space model dimension is $d_n=(n+1)^4$.}
	\end{figure}
	
	The state $\varrho_\textsc{SPDC}$ may directly be taken as the source emitting photons in the two spatial modes~$a$ and~$b$. In this case, it is desirable to pick a small squeezing strength to increase the probability of single-photon events. Here, we set $\tau\approx0.2418\equiv 2.1~\mathrm{dB}$ so that this probability is approximately 5\%. With a detector efficiency of $\eta=0.9$ and a presumed PNRD resolution of $n_0=3$, the total POVM for the two spatial modes is defined as $\{\Pi^{(\eta,a)}_{n_\textsc{h},n_\textsc{v}}(\vartheta_1,\varphi_1)\otimes\Pi^{(\eta,b)}_{m_\textsc{h},m_\textsc{v}}(\vartheta_2,\varphi_2)\}$, where the photon numbers each runs from~0 to~3 and the angle pairs are assigned from a list of 7 settings---corresponding to $L_\mathrm{pol}=44$ linearly independent outcomes per spatial mode---that would give a total of 49 measurement settings for the two spatial modes. These are sufficient to uniquely determine the polarization sector of~$\varrho$.
	
	Figure~\ref{fig:two-mode-source} shows that the TMSV state~$\varrho_\textsc{SPDC}$ can be described by polarization Fock kets of the effective dimension $\dRB=3$. This implies, with extremely high evidence strength, that the source is emitting $n_\mathrm{RB}=2$~photons \emph{per spatial mode}. 
	
	\section{Parametric model certification with the relative belief}
	\label{sec:decoherence}
	
	The number of degrees of freedom of a physical system that can be measured precisely or modeled effectively is limited due to practical feasibility. For instance, one cannot hope to describe and measure all microscopic physical properties of an Avogadro number of molecules. Typically, only systems with degrees of freedom that are orders of magnitude smaller are treatable with precise models. The two-level atom or a mode of light are two such very simple systems for which the quantum state can be explicitly stated.
	
	In this section, we discuss the possibility for RB to ascertain the number of \emph{external} degrees of freedom that are involved when the physical system an observer has access to interacts with some an external environment that is not measured, \emph{provided} that information about the latter is encoded in the reduced state of the physical system. For this purpose, the number of external degrees of freedom cannot be too large to allow for a precise modeling of the actual interaction process. Different interaction models may then be put to the test based on the data obtained from measuring the physical system, and the usual elements of RB certification shall decide which models are plausible.
	
	More specifically, we investigate two different types of interaction models that are simplified cases of actual physical situations and may be universally described by variants of the Jaynes--Cummings model~\cite{Vivek:2023nonequilibrium,Cheng:2023quantum,Cui:2023effective,Nodurft:2019optical,Ezaki:1995photon,Seke:1985extended,Jaynes:1963comparison}. They are the atom-field and field-absorber interaction~models. We first discuss the negative case where RB fails to distinguish between different numbers of interacting photon fields for the atom-field interaction scenario. After which, in the positive case where a single-mode photon field interacts with multiple molecular absorbers, we demonstrate the capability of RB in distinguishing the number of these absorbers through data collected from the photon field after tracing over the absorber degrees of freedom.
	
	\subsection{Atom-field interaction (negative case)}
	\label{subsec:atom-field}
	
	Consider a two-level atom that is initially prepared in some state $\varrho_0$. This atom undergoes an interaction process with $n_\mathrm{field}$ independent photon-field modes that is described by the interaction Hamilton operator
	\begin{equation}
		H_\mathrm{int}=\sum^{n_\mathrm{field}}_{j=1}g_j\left(\sigma_- a^\dag_j+\sigma_+ a_j\right)\,,
		\label{eq:Hint_AF}
	\end{equation}
	where $g_0=0$, $\sigma_+=\ket{\mathrm{e}}\bra{\mathrm{g}}=\sigma_-^\dag$, and both $\ket{\mathrm{g}}\bra{\mathrm{g}}$ and $\ket{\mathrm{e}}\bra{\mathrm{e}}$ denote the atomic ground and excited states respectively. This form generalizes the original Jaynes--Cummings model~\cite{Jaynes:1963comparison} to more than one field mode. Equation~\eqref{eq:Hint_AF} then corresponds to the interaction unitary operator $U_\mathrm{int}=\E{-\I H_{\mathrm{int}}}$, where we shall henceforth set the evolution time $t\equiv1$ for convenience, such that the reduced state $\varrho(n_\mathrm{field},\rvec{g}_{n_\mathrm{field}})=\ptr{\mathrm{field}}{U_\mathrm{int}\varrho_0\otimes\vacket \vacbra U_\mathrm{int}^\dag}$ now represents the atom-field interaction model defined by the row of coupling strengths $\rvec{g}_{k}=(g_1\,\,g_2\,\,\ldots\,\,g_k)$ for some integer $k>0$. Note that $\varrho(0,0)=\varrho_0$.
	
	\begin{figure}[t]
		\includegraphics[width=1\columnwidth]{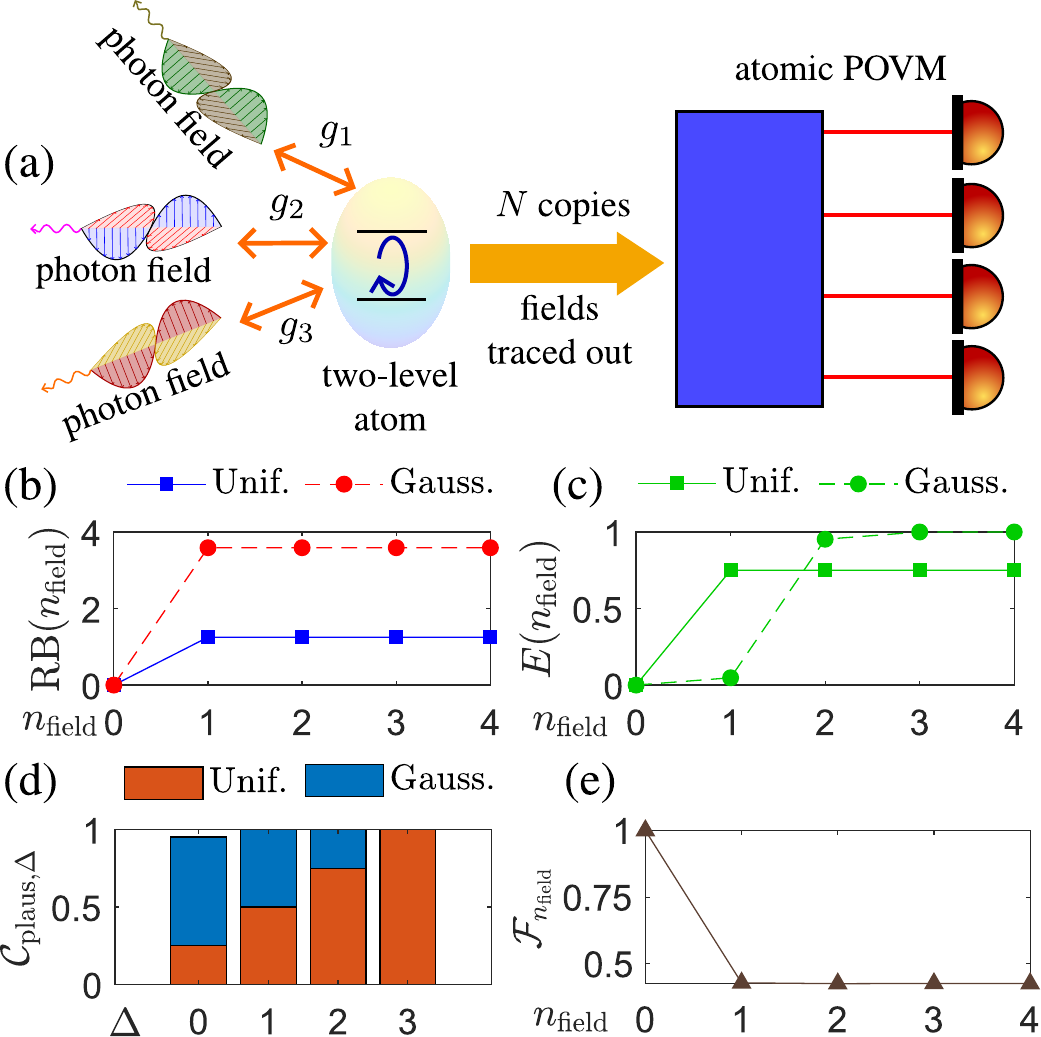}
		\caption{\label{fig:AD}Atomic decoherence-model certification with~RB, where (a)~an atom prepared in the superposition state ket $(\ket{\mathrm{g}}+\ket{\mathrm{e}})/\sqrt{2}$ is coupled to three photon-field modes of strengths $g_1=g_2=g_3=1$. Multiple copies of the interacting atom~($N=1000$) are measured (with all field modes traced out) using a fixed randomly-chosen atom POVM consisting of four rank-one outcomes. The (b)~RB ratios, (c)~evidence strengths and (d)~plausible-interval credibilities for both the uniform and Gaussian~[$\propto \exp(-n_\mathrm{field}^2)$] prior distributions are~plotted. (e)~The fidelity with the ideal superposition atomic state saturates to the correct value that is significantly less than~1. The incapability of RB to tell the number of interacting field modes is clear in this figure.}
	\end{figure}
	
	Suppose that the \emph{true} model is $\varrho=\varrho(3,\rvec{g}_3)$ with a pure initial atomic state $\varrho_0=\ket{\,\,\,}\bra{\,\,\,}$ and  $\ket{\,\,\,}=(\ket{\mathrm{g}}+\ket{\mathrm{e}})/\sqrt{2}$, where $\rvec{g}_{3}=(1\,\,\,0.5\,\,\,0.1)$ signifies uneven coupling with the three photon-field modes. The observer is ignorant about the true coupling strengths \emph{and} $n_\mathrm{field}$, but believes that $\varrho(n_\mathrm{field},\rvec{g}_{n_\mathrm{field}})$ are viable models that represent the \emph{dominant} environmental interaction on the \emph{expected (known) initially-prepared atomic state} $\varrho_0$. Then, the procedure of RB model certification proceeds as follows:
	
	\begin{enumerate}
		\item Decide on a model set of finite cardinality $K$, $\{\varrho(0,0),\varrho(1,g_1),\varrho(2,\rvec{g}_2),\ldots,\varrho(K,\rvec{g}_K)\}$, to be tested with~RB.
		\item Assign a prior probability $\PR(n_\mathrm{field})$ to each model for $0\leq n_\mathrm{field}\leq K$.
		\item Perform a measurement on the unknown atomic state using a POVM and obtain a dataset $\mathbb{D}$.
		\item Starting with $n_\mathrm{field}=0$:
		\begin{itemize}
			\item[(i)] Compute $L_{n_\mathrm{field}}=\displaystyle\max_{\rvec{g}_{n_\mathrm{field}}}\{L(\mathbb{D}|\varrho(n_\mathrm{field},\rvec{g}_{n_\mathrm{field}}))\}$ and store this value. 
			\item[(ii)] Increase $n_\mathrm{field}$ by 1.
			\item[(iii)] Repeat (i)--(iii) for all $0\leq n_\mathrm{field}\leq K$.
		\end{itemize}
		\item Compute and store the posterior probabilities $\PR(n_\mathrm{field}|\mathbb{D})$ for $0\leq n_\mathrm{field}\leq K$.
		\item Compute $\RB(n_\mathrm{field})$ for all $0\leq n_\mathrm{field}\leq K$ and find the smallest plausible $n_\mathrm{field}$. All larger values will also be plausible.
		\item Compute evidence strengths $E(n_\mathrm{field})$ for all $0\leq n_\mathrm{field}\leq K$, and~$\mathcal{C}_{\mathrm{plaus},\Delta}$.
	\end{enumerate}
	
	Figure~\ref{fig:AD} illustrates the performance of RB model certification for $g_1=g_2=g_3=g=1$. In this scenario, RB always predicts only \emph{one} photon-field mode is interacting with the atom ($n_\mathrm{field}\geq1$) with high evidence strength. One can understand why this is so by defining the ladder operator $A=\sum^{n_\mathrm{field}}_{j=1}g_ja_j/\sqrt{\sum^{n_\mathrm{field}}_{j'=1}g_{j'}^2}$ such that $[A,A^\dag]=1$ and rewrite the \emph{same} interaction Hamilton operator in Eq.~\eqref{eq:Hint_AF} as
	\begin{equation}
		H_\mathrm{int}=g'\left(\sigma_- A^\dag+\sigma_+ A\right)\,,
		\label{eq:Hint_AF2}
	\end{equation}
	with $g'=\sqrt{\sum^{n_\mathrm{field}}_{j=1}g_{j}^2}$, which is equivalent to that for a two-level atom interacting with a single photon-field mode of an effective coupling strength~$g'$. Therefore, the ML procedure simply finds the most likely~$g$ for this Hamilton operator. Indeed, for the dataset $\mathbb{D}$ used in Fig.~\ref{fig:AD}, the ML estimators for~$g'$ in the range $1\leq n_\mathrm{field}\leq4$ are respectively 1.7143, 1.7206, 1.7191 and 1.7196, which are close to $\sqrt{3}\approx1.7321$ and all corresponding to the same maximum log-likelihood of~$-1202.6$.
	
	\subsection{Field-absorber interaction (positive case)}
	\label{subsec:field-abs}
	
	We may conceive of another scenario that extends the Jaynes--Cummings model to multiple interacting molecules instead. Let us consider a single-mode photon field traversing through an ensemble of two-level molecules that absorb photons, which are all initially in the ground state~$\ket{\mathrm{g}}\bra{\mathrm{g}}$. The Hamilton operator describing the interaction between the single-mode field and $n_\mathrm{abs}$ molecular absorbers may be written as
	\begin{equation}
		H_\mathrm{int}=\sum^{n_\mathrm{abs}}_{j=1}g_j\left(\sigma_{-,j} a^\dag+\sigma_{+,j} a\right)\,,
		\label{eq:Hint_FA}
	\end{equation}
	where $g_0=0$ and $\sigma_{\pm,j}$ is $\sigma_\pm$ for the $j$th absorber. This is in fact the Tavis--Cummings model~\cite{Tavis:1968exact,Tavis:1969approximate} under the rotating-wave approximation. Several other models of this kind include the $n_\mathrm{abs}=1$ Rabi model~\cite{Rabi:1936process,Braak:2011integrability} and the more general Dicke model~\cite{Dicke:1954coherence} when off-resonant terms are included. These models helped establish the foundation of superradiance~\cite{Dicke:1954coherence,Gross:1982superradiance,Scully:2009super,Mlynek:2014observation,Nefedkin:2017nonlinear,Masson:2022universality} in quantum optics.
	
	Similar to Sec.~\ref{subsec:atom-field}, the corresponding model as a result of the interaction unitary $U_\mathrm{int}=\E{-\I H_\mathrm{int}}$ ($t\equiv1$) is the reduced optical state $\varrho(n_\mathrm{abs},\rvec{g}_{n_\mathrm{abs}})=\ptr{\mathrm{abs}}{U_\mathrm{int}\varrho_0\otimes\left(\ket{\mathrm{g}}\bra{\mathrm{g}}\right)^{\otimes n_\mathrm{abs}}U^\dag_{\mathrm{int}}}$, with $\varrho_0=\varrho(0,0)$ being the initially-prepared single-mode field state and $\rvec{g}_{k}=(g_1\,\,g_2\,\,\ldots\,\,g_k)$ for some integer $k>0$.
	
	\begin{figure}[t]
		\includegraphics[width=1\columnwidth]{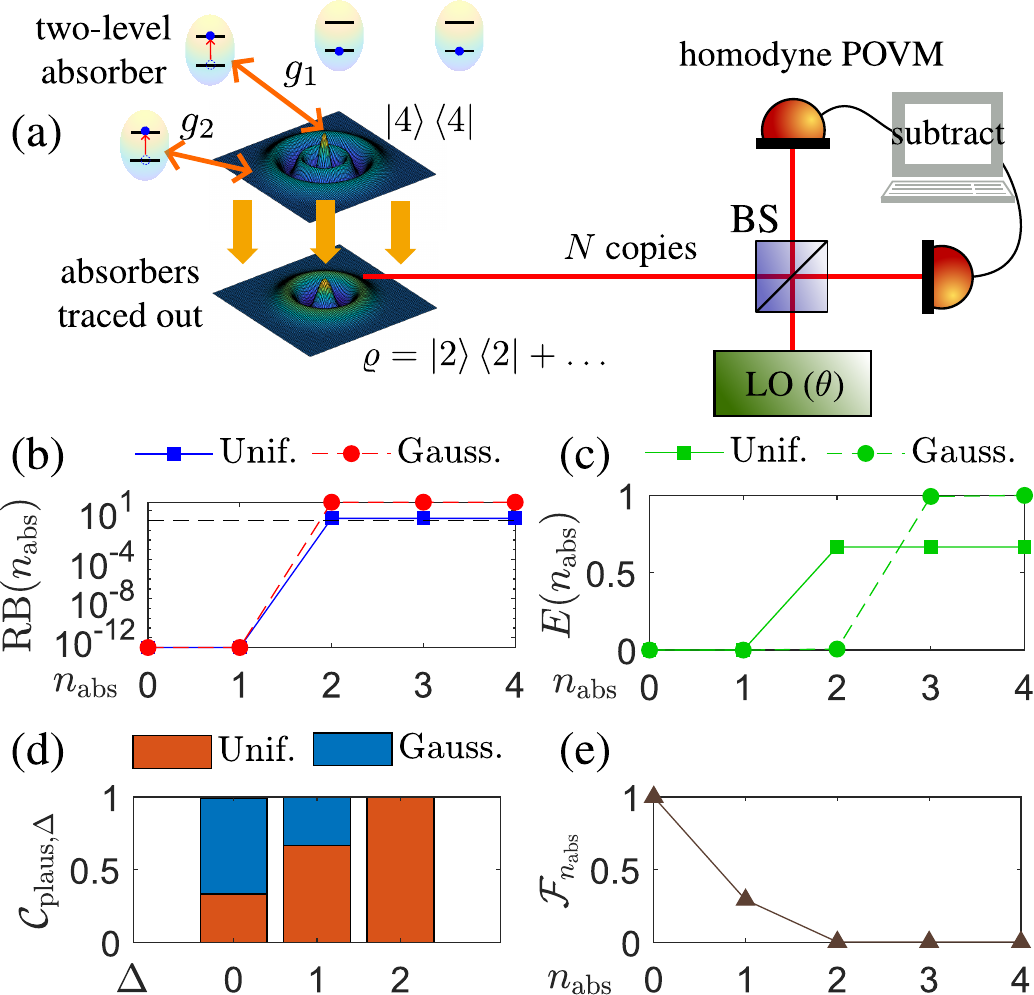}
		\caption{\label{fig:loss}Optical absorption model certification with~RB, where (a)~the Fock state $\ket{4}\bra{4}$ couples to two absorber molecules with equal strengths ($g_1=1=g_2$). Multiple copies of the resulting optical state are then sent to a standard homodyne setup, where the phase~$\theta$ of the local oscillator~(LO) determines the \mbox{quadrature angle}. Both the Hilbert space of $\varrho$ and number of angles are set to $d=9$ and 10. A total of $N=10^6$ copies are measured using all these settings. The (b)~RB ratios (all greater than~1 for $n_\mathrm{abs}>1$), (c)~evidence strengths and (d)~plausible-interval credibilities for both the uniform and Gaussian~[$\propto \exp(-n_\mathrm{abs}^2)$] prior distributions are~plotted. (e)~The fidelity with the ideal Fock state drops gradually to almost zero as RB approaches the correct number of absorbers $n_\mathrm{abs}$, since the actual mixed state is given by $\varrho\,\,\widehat{=}\,\,\mathrm{diag}(0,0,0.8159,0.1822,0.0019,0,0,0,0)\cong\ket{2}\bra{2}$, which is almost completely orthonormal to $\ket{4}\bra{4}$.}
	\end{figure}
	
	As an example, we take the \emph{true} model to be $\varrho=\varrho(2,\rvec{g}_2)$ with the four-photon Fock state $\varrho_0=\ket{4}\bra{4}$ as the initially-prepared single-mode photon-field state, where $\rvec{g}_{2}=(1\,\,\,1)$. Again, the observer is ignorant about the true coupling strengths \emph{and} $n_\mathrm{abs}$, but thinks that $\varrho(n_\mathrm{abs},\rvec{g}_{n_\mathrm{abs}})$ are the possible models governing the \emph{dominant} interaction type on the \emph{expected (known) initial optical state}, which is $\ket{4}\bra{4}$ in \mbox{this~example}. 
	
	To illustrate the performance of RB model certification with a widely-known measurement scheme, we employ the homodyne POVM~\cite{Yuen:1983ba,Abbas:1983ak, Schumaker:1984qm} consisting of the projector outcomes $\E{\I\theta a^\dag a}\ket{x}\bra{x}\E{-\I\theta a^\dag a}$, where $\ket{x}\bra{x}$ are the quadrature eigenstates and the unitary operator $\E{\I\theta a^\dag a}$ rotates these eigenstates in phase space to generate a set of informationally complete~POVM. For a truncated Fock space of dimension~$d$, it is known that $d$ different values of $\theta$ give rise to an informationally complete POVM that can uniquely reconstruct the optical state in this truncated space~\cite{Sych:2012informational}. In order to ascertain the minimum possible value of~$d$ for a faithful description of the single-mode optical state~$\varrho$, one should carry out RBDC as described in Secs.~\ref{subsec:RBDC} and \ref{sec:imp_sources} using the homodyne POVM, so that no assumptions are left uncertified.
	
	Suppose, after RBDC, the effective dimension of~$\varrho$ is $\dRB$. Then, the RB model certification route for testing the different values of $n_\mathrm{abs}$ is highly similar to that presented in Sec.~\ref{subsec:atom-field}:
	
	\begin{enumerate}
		\item Decide on a model set of finite cardinality $K$, $\{\varrho(0,0),\varrho(1,g_1),\varrho(2,\rvec{g}_2),\ldots,\varrho(K,\rvec{g}_K)\}$, to be tested with~RB.
		\item Assign a prior probability $\PR(n_\mathrm{abs})$ to each model for $0\leq n_\mathrm{abs}\leq K$.
		\item Starting with $n_\mathrm{abs}=0$:
		\begin{itemize}
			\item[(i)] With the homodyne-POVM dataset $\mathbb{D}$ and $\dRB$ found in RBDC, compute and store $L_{n_\mathrm{abs}}=\displaystyle\max_{\rvec{g}_{n_\mathrm{abs}}}\{L(\mathbb{D}|\varrho(n_\mathrm{abs},\rvec{g}_{n_\mathrm{abs}}))\}$ for all $\dRB$-dimensional $\varrho(n_\mathrm{abs},\rvec{g}_{n_\mathrm{abs}})$. 
			\item[(ii)] Increase $n_\mathrm{abs}$ by 1.
			\item[(iii)] Repeat (i)--(iii) for all $0\leq n_\mathrm{abs}\leq K$.
		\end{itemize}
		\item Compute and store the posterior probabilities $\PR(n_\mathrm{abs}|\mathbb{D})$ for $0\leq n_\mathrm{abs}\leq K$.
		\item Compute $\RB(n_\mathrm{abs})$ for all $0\leq n_\mathrm{abs}\leq K$ and find the smallest plausible $n_\mathrm{abs}$. All larger values will also be~plausible.
		\item Compute evidence strengths $E(n_\mathrm{abs})$ for all $0\leq n_\mathrm{abs}\leq K$, and~$\mathcal{C}_{\mathrm{plaus},\Delta}$.
	\end{enumerate}
	
	This time, owing to the distinguishable differences in the operator  $\sum^{n_\mathrm{abs}}_{j=1}g_j\sigma_{-,j}$ for various $n_\mathrm{abs}$ in the Tavis--Cummings model, Fig.~\ref{fig:loss} showcases the capability of RB to discern the plausible optical absorption models from the implausible ones. With respect to these optical-absorption interaction models, RB predicts that the plausible numbers of interacting absorbers are $n_\mathrm{abs}\geq2$ with very high evidence strength based on the obtained dataset for a total of $N=10^6$ sampling~copies.
	
	To see why this model brings about distinguishable interacting degrees of freedom, we consider the case for which $g_j=g$ without loss of generality, so that the operator $J_-=\sum^{n_\mathrm{abs}}_{j=1}\sigma_{-,j}$ may then be understood as the $n_\mathrm{abs}$-spin lowering operator that accesses the collective subspace spanned by the angular-momentum eigenstates $\ket{j,m}$, where $|m|\leq j\leq n_\mathrm{abs}/2$ and $m=(n_\mathrm{e}-n_\mathrm{g})/2$ is half the difference between the population numbers in the excited and ground states that takes values $m\in\{-j,-j+1,-j+2,...\,,j-2,j-1,j\}$~\cite{Nussenzveig:1973introduction}. This accessible subspace by $J_-$ is directly dependent on $n_\mathrm{abs}$ and this feature is nontrivially transferred to the photonic state after tracing over the absorber degrees of~freedom.
	
	\section{Conclusion}
	
	In this work, we formally introduce a powerful Bayesian-inference paradigm for quantum information theory that utilizes the mechanism of relative-belief reasoning to correctly evaluate a bounded set of hypotheses about a given quantum system and decide which of these hypotheses are plausible. Such an evidence-based decision making procedure solely relies on the measurement data to compare all incorporated initial beliefs, expressed in the form of prior probabilities, about these hypotheses against the corresponding posterior probabilities that measures our beliefs after the experiment. This is accompanied by the built-in technical capabilities to quantify the evidence strength for all plausible hypotheses and assign Bayesian credibilities of plausible intervals. Recent advancement in high-precision storage numerical formats helps to turn this relative-belief paradigm into an operational reality.
	
	As a procedure for choosing Hilbert spaces of plausible dimensions to describe an unknown quantum state, we showed that the relative-belief dimension certification protocol is capable of selecting plausible dimensions from the data such that the smallest plausible dimension is, typically, never smaller than those chosen by optimizing a broad class of information criteria, which include Akaike's and Bayesian information criteria. This firmly establishes the relative-belief methodology as the more conservative paradigm for evaluating dimension-valued~hypotheses.
	
	As part of a crucial step in state preparation for quantum-information and communication tasks, we demonstrate how relative belief can be feasibly used to evaluate the quality of quantum sources that were claimed to emit a certain number of photons. We show for spatial-mode sources, the relative-belief dimension certification is able to reliably arrest any spurious additional photons generated by the source and report the correct plausible Hilbert spaces needed to faithfully describe the true state.
	
	We go beyond dimension certification and discuss more general model certification with relative belief. In particular, we answer the question of whether a purely evidence-based Bayesian reasoning can assist in ascertaining the number of external degrees of freedom in an interacting environment with an accessible physical system without directly measuring and obtaining explicit information about the external environment. We found that under the premise that information concerning the interacting environment is carried over to the measured physical system after tracing over the external degrees of freedom and that these degrees of freedom are sufficiently small to be precisely accounted for using mathematical models, the relative-belief methodology can certify which interaction models out of a finite set most plausibly describe the given system-environment interaction. This is illustrated with the Tavis--Cummings model describing the interaction between a single-mode photon field and molecular absorbers that carries nontrivial dependence on the number of absorbers that is transferred to the physical system even after tracing over the absorber degrees of freedom.
	
	The interesting results in this exposition are a testament to the role of the relative-belief inference methodology not just as a model-dimension selection rule, but also as a general hypotheses certification protocol that is potentially pivotal in addressing foundational issues in quantum mechanics. The versatile and operational features of this methodology makes the concept of relative belief applicable to arguably any inference problems in quantum information theory, and it is our belief that deeper explorations into these uncharted avenues shall bear invaluable fruits of~labor.

	\begin{acknowledgments}
		The authors thank J.~Sperling and J.~Gil-Lopez for insightful discussions. This work was supported by the European Union’s Horizon 2020 Research and Innovation Programme Grant No. 899587 (Project Stormytune). Y.S.T., S.U.S. and H.J. acknowledge support from the National Research Foundation of Korea (NRF) grants funded by the Korean government (Grant Nos. NRF2020R1A2C1008609, 2023R1A2C1006115, NRF2022M3E4A1076099 and RS-2023-00237959) \emph{via} the Institute of Applied Physics at Seoul National University, the Institute of Information \& Communications Technology Planning \& Evaluation (IITP) grant funded by the Korea government (MSIT) (IITP-2021-0-01059 and IITP-20232020-0-01606), and the Brain Korea 21 FOUR Project grant funded by the Korean Ministry of Education. M.E. was supported by a grant form the Natural Sciences and Engineering Research Council of Canada 2017-06758. L.L.S.S.  acknowledges support from Ministerio de Ciencia e Innovaci\'on (Grant PID2021-127781NB-I00).
	\end{acknowledgments}
	
	\appendix
	
	\section{Relative-belief formalism in the continuous case}
	\label{app:RBcont}
	
	For a bounded continuous set of hypotheses $\{\mathcal{H}(\rvec{x})\}$, then in terms of the prior density function $\PR(\rvec{x})$, the posterior density function is given by
	\begin{equation}
		\PR(\rvec{x}|\mathbb{D})=\dfrac{L(\mathbb{D}|\rvec{x})\PR(\rvec{x})}{\int(\D\rvec{x}')L(\mathbb{D}|\rvec{x}')\PR(\rvec{x}')}\,,
	\end{equation}
	where $(\D\rvec{x}')$ is some volume measure for the integration. Consequently, the continuous RB ratio in $\rvec{x}$ is given~by
	\begin{equation}
		\RB(\rvec{x})=\dfrac{\PR(\rvec{x}|\mathbb{D})}{\PR(\rvec{x})}=\dfrac{L(\mathbb{D}|\rvec{x})}{\int(\D\rvec{x}')L(\mathbb{D}|\rvec{x}')\PR(\rvec{x}')}\,.
		\label{eq:RBratio_cont}
	\end{equation} 
	The evidence strength straightforwardly modifies~to
	\begin{equation}
		E(\rvec{x})=\int(\D\rvec{x}')\PR(\RB(\rvec{x}')\leq\RB(\rvec{x})|\mathbb{D})\,,
	\end{equation}
	where the interpretations remain unchanged as in Fig.~\ref{fig:Ek}. 
	
	If $\mathcal{P}$ is some parameter subspace containing $\rvec{x}_\mathrm{eff}$, a hypothesis corresponding to the smallest likelihood for which $\RB(\rvec{x}_\mathrm{eff})>1$, then analogous plausible regions and their prior and posterior contents may be defined as
	\begin{align}
		\mathcal{R}_{\mathrm{plaus},\mathcal{P}}=&\,\mathcal{P}\,,\nonumber\\
		\mathcal{S}_{\mathrm{plaus},\mathcal{P}}=&\,\int_\mathcal{P}(\D\rvec{x})\PR(\rvec{x})\,,\nonumber\\
		\mathcal{C}_{\mathrm{plaus},\mathcal{P}}=&\,\int_\mathcal{P}(\D\rvec{x})\PR(\rvec{x}|\mathbb{D})>\mathcal{S}_{\mathrm{plaus},\mathcal{P}}\,.
	\end{align}
	Note that in the continuous case, it is often more favorable to define a plausible region that is convex. In this case, if $\log L(\mathbb{D}|\rvec{x})$ is concave in $\rvec{x}$, then the entire space $\mathcal{P}=\{\rvec{x}'|\RB(\rvec{x}')>1\}$ is to be taken, since all elements are closed under the convex sum. Finally, the type-I and type-II error probabilities are
	\begin{align}
		\epsilon_\mathrm{I}=\int_{\mathcal{T}}(\D\rvec{x})\,\PR(\rvec{x})\,\eta(\RB(\rvec{x})-1)\,,\nonumber\\
		\epsilon_\mathrm{II}=\int_{\mathcal{T}^{\mathrm{c}}}(\D\rvec{x})\,\PR(\rvec{x})\,\eta(1-\RB(\rvec{x}))\,.
	\end{align}
	
	\section{The full RBDC scheme for any data sample size}
	\label{app:fullRBDC}
	
	The start of the most holistic methodology of RBDC is to assign prior probabilities $\PR_1(d_k)$ and $\PR_2(\varrho(d_k))$ to \emph{both} the dimension hypothesis set and states in the quantum state space~$\mathcal{Q}_{d_{k}}$ for the $d_k$-dimensional Hilbert space, with $\sum^K_{k=1}\PR_1(d_k)=1=\int_{\mathcal{Q}_{d_{k}}}(\D\varrho(d_{k}))\,\PR_2(\varrho(d_k))$ and $(\D\varrho(d_{k}))$ is the volume measure for $\mathcal{Q}_{d_{k}}$. The product of these prior probabilities $\PR(d_k,\varrho(d_k))=\PR_1(d_k)\,\PR_2(\varrho(d_k))$ is then the \emph{prior} for RBDC. The posterior probability is then given by
	\begin{align}
		&\,\PR(d_k,\varrho(d_k)|\mathbb{D})\nonumber\\
		=&\,\dfrac{\PR(d_k,\varrho(d_k))\,L(\mathbb{D}|\varrho(d_k))}{\sum^K_{k'=1}\int_{\mathcal{Q}_{d_{k'}}}(\D\varrho(d_{k'}))\,\PR(d_{k'},\varrho(d_{k'}))\,L(\mathbb{D}|\varrho(d_{k'}))}\,.
	\end{align}
	
	Since the dimension hypothesis is of interest here, we may consider the marginal posterior distribution of $d_k$, where marginalization is performed over \emph{all} state parameters:
	\begin{equation}
		\PR(d_k|\mathbb{D})=\int_{\mathcal{Q}_{d_k}}(\D\varrho(d_k))\,\PR(d_k,\varrho(d_k)|\mathbb{D})\,.
	\end{equation}
	
	The corresponding RB ratio reads
	\begin{align}
		&\,\RB(d_k)=\dfrac{\PR(d_k|\mathbb{D})}{\PR_1(d_k)}\nonumber\\
		=&\,\dfrac{\int_{\mathcal{Q}_{d_{k}}}(\D\varrho(d_{k}))\,\PR_2(\varrho(d_{k}))\,L(\mathbb{D}|\varrho(d_{k}))}{\sum^K_{k'=1}\int_{\mathcal{Q}_{d_{k'}}}(\D\varrho'(d_{k'}))\,\PR(d_{k'},\varrho'(d_{k'}))\,L(\mathbb{D}|\varrho'(d_{k'}))}\,.
		\label{eq:RBratio_holistic}
	\end{align}
	
	Hence, the main differences between this more general RBDC scheme and that described in Sec.~\ref{subsec:RBDC} is the replacement of the simple likelihood maximization over all $\varrho(d_k)$ for each $d_k$ in the posterior computations with more sophisticated state averages and the use of the RB ratio in~\eqref{eq:RBratio_holistic}. 
	
	For large datasets that need not correspond to an astronomical $N$, the likelihood $L(\mathbb{D}|\varrho(d_{k}))$ would peak strongly around $\ML(d_k)$, the maximum in~$\mathcal{Q}_{d_k}$, such that
	\begin{align}
		&\,\int_{\mathcal{Q}_{d_{k}}}(\D\varrho(d_{k}))\,\PR(d_{k'},\varrho'(d_{k'}))\,L(\mathbb{D}|\varrho(d_{k}))\nonumber\\
		\cong&\,\PR_1(d_{k})\,L(\mathbb{D}|\ML(d_{k}))\int_{\mathcal{Q}_{d_{k}}}(\D\varrho(d_{k}))\,\PR_2(\varrho(d_{k}))\nonumber\\
		=&\,\PR_1(d_{k})\,L(\mathbb{D}|\ML(d_{k}))
	\end{align}
	and
	\begin{equation}
		\RB(d_k)\rightarrow\dfrac{L(\mathbb{D}|\ML(d_{k}))}{\sum^K_{k'=1}\PR_1(d_{k'})\,L(\mathbb{D}|\ML(d_{k'}))}\,.
	\end{equation}
	Therefore, both RBDC schemes asymptotically coincide. For highly data-limited situations, where $L(\mathbb{D}|\varrho(d_{k}))$ is a smooth broad function over $\mathcal{Q}_{d_k}$, then there is no way around the state-space integrals other than computing them explicitly. This requires highly sophisticated numerical Monte~Carlo techniques~\cite{Shang:2015mc,Seah:2015mc} that pay close attention to the boundaries of $\mathcal{Q}_{d_k}$ that generally lack a complete analytical understanding owing to their complexity. Such discussions are beyond the scope of this article.
	
	\section{Linear independence of the polarimetric POVM}
	\label{app:linindep}
	
	An important observation in \cite{Goldberg:2021quantum} that \emph{any} quantum-polarimetry measurement can only probe the \emph{polarization sector} of a quantum state $\varrho$. To define this sector, we rewrite the two-mode Fock kets $\ket{n_1,n_2}$ as a spin angular-momentum eigenstate $\ket{S,m}$, where $S=(n_1+n_2)/2$ and $m=(n_1-n_2)/2$. Then, the polarization sector (or polarization state) $\varrho_\mathrm{pol}$ is defined as
	\begin{align}
		\varrho_\mathrm{pol}=&\,\bigoplus^\infty_{S=0}p(S)\varrho^{(S)}\nonumber\\
		=&\,\sum^\infty_{S=0}p(S)\sum^{S}_{m=-S}\sum^{S}_{m'=-S}\ket{S,m}\varrho^{(S)}_{m,m'}\bra{S,m'}\,.
	\end{align}
	In other words, in the $\{\ket{S,m}\}$ basis, \emph{only} constant-$S$ sectors~$\varrho^{(S)}$ of $\varrho$ are probed by polarimetric POVMs.
	
	\begin{figure}[b]
		\includegraphics[width=1\columnwidth]{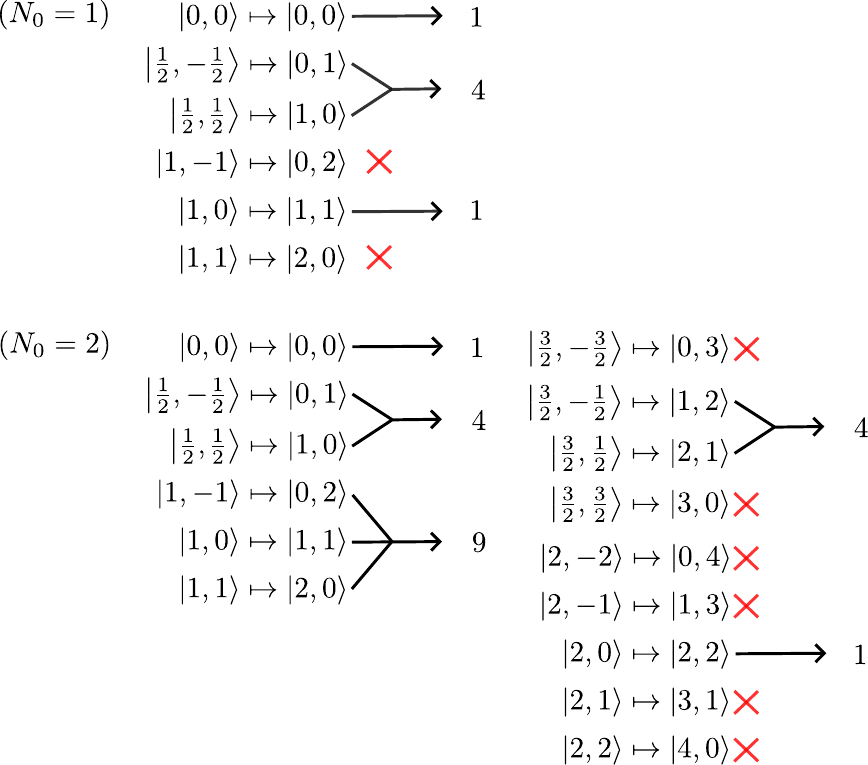}
		\caption{\label{fig:N1_N2}Equivalence between number and angular-momentum eigenstates and the appropriate subspaces probed by quantum polarimetry.}
	\end{figure}
	
	If $n_0$ is now the largest photon number detectable for \emph{each} polarization arm, which is our experimental scenario, then the following counting scheme gives the correct $M_\mathrm{pol}$. Supposing that $n_0=1$, then the polarimetric POVM $\{\Pi^{(\eta)}_{n_1,n_2}(\vartheta,\varphi)\}$ can probe sectors up to $S=1$. The relevant polarization sector from the maps
	would then contain a one-dimensional vacuum sector, a two-dimensional one-photon Hermitian sector that gives 4 real parameters, and a one-dimensional two-photon sector. This tallies to $6$ free parameters. Notice that $\ket{0,2}$ and $\ket{2,0}$ are excluded since $n_0=1$ at each port. Thus, $M_\mathrm{pol}=1+4+1=6$ when $n_0=1$. For $n_0=2$, the POVM can probe sectors up to $S=2$, so that
	implies that $M_\mathrm{pol}=1+4+9+4+1=19$. From the above patterns, it is easy to deduce Eq.~\eqref{eq:Mpol}.

\end{document}